\documentclass[envcountsame]{llncs}

\usepackage{times}
\usepackage{url}
\usepackage{cite}
\usepackage{amsmath}
\usepackage{txfonts}
\usepackage{wrapfig}

\usepackage{cite}
\usepackage{amsfonts}
\usepackage{amssymb}
\usepackage{amstext}
\usepackage{amsmath}
\usepackage{mathtools}
\usepackage{paralist}

\usepackage{tikz}
\usetikzlibrary{calc}
\usetikzlibrary{shapes,arrows,positioning,shadows,snakes}

\usepackage{footmisc}

\renewenvironment{proof}{\noindent{\it{Proof}}: }   {\hfill\qed\smallskip\par}

\usepackage{pifont}
\usepackage[hang,small,bf]{caption}
\usepackage{subcaption}

\newcommand{\pred}{ {\mathit{pred}}}
\newcommand{\suc}{\mathit {succ}}
\newcommand{\child}{\mathit{child}}

\usepackage{nameref}
\usepackage[linktocpage=true,pagebackref=true]{hyperref}
\usepackage{cleveref}

\usepackage{thmtools,thm-restate} 

\newtheorem{fact}[theorem]{Fact}
\newcommand{\greedy}{{\sc Greedy\text{ }BST}\xspace} 
\newcommand{\greedyfuture}{{\sc GreedyFuture}\xspace} 

\usepackage{graphicx}
\usepackage{graphics}
\usepackage{colordvi}
\usepackage{xspace}
\usepackage{algorithm}
\usepackage{algorithmicx}
\usepackage{algpseudocode}
\usepackage{url}
\usepackage{enumitem}
\usepackage{ wasysym }
\usepackage{ textcomp }

\def\ShowComment{True}

\ifdefined\ShowComment

\def\thatchaphol#1{\marginpar{$\leftarrow$\fbox{T}}\footnote{$\Rightarrow$~{\sf #1 --Thatchaphol}}}

\else

\def\thatchaphol#1{}

\fi

\newcommand{\LK}[1]{\textcolor{blue}{LK: #1}}
\newcommand{\stair}{{\mathit{stair}}}

\title{\vspace{-1ex} Greedy Is an Almost Optimal Deque}

\date{}

\author{
Parinya~Chalermsook\inst{1},
Mayank~Goswami\inst{1},
L\'{a}szl\'{o}~Kozma\inst{2},
Kurt~Mehlhorn\inst{1},
and
Thatchaphol~Saranurak\inst{3}\thanks{Work mostly done while at Saarland University.}
}

\institute{Max-Planck Institute for Informatics, Saarbr\"{u}cken, Germany 66123\\ 
\and Department of Computer Science, Saarland University, Saarbr\"{u}cken, Germany 66123\\ 
\and KTH Royal Institute of Technology, Stockholm, Sweden 11428\\ 
}

\begin{document}
\pagenumbering{arabic}

\maketitle

\vspace{-2em}

\begin{abstract}
In this paper we extend the geometric binary search tree (BST) model of Demaine, Harmon, Iacono, Kane, and P\v{a}tra\c{s}cu (DHIKP) to accommodate for insertions and deletions. Within this extended model, we study the online \greedy algorithm introduced by DHIKP. \ \greedy is known to be equivalent to a maximally greedy (but inherently offline) algorithm introduced independently by Lucas in 1988 and Munro in 2000, conjectured to be dynamically optimal. 

With the application of forbidden-submatrix theory, we prove a quasilinear upper bound on the performance of \greedy on deque sequences. 
It has been conjectured (Tarjan, 1985) that splay trees (Sleator and Tarjan, 1983) can serve such sequences in linear time. Currently neither splay trees, nor other general-purpose BST algorithms are known to fulfill this requirement. As a special case, we show that \greedy can serve output-restricted deque sequences in linear time. A similar result is known for splay trees (Tarjan, 1985; Elmasry, 2004).

As a further application of the insert-delete model, we give a simple proof that, given a set $U$ of permutations of $[n]$, the access cost of any BST algorithm is $\Omega( \log \vert U \vert + n)$ on ``most'' of the permutations from $U$. In particular, this implies that the access cost for a random permutation of $[n]$ is $\Omega{(n\log{n})}$ with high probability. 

Besides the splay tree noted before, \greedy has recently emerged as a plausible candidate for dynamic optimality. Compared to splay trees, much less effort has gone into analyzing \greedy. Our work is intended as a step towards a full understanding of \greedy, and we remark that forbidden-submatrix arguments seem particularly well suited for carrying out this program.

\end{abstract}

\section{Introduction}

Binary search trees (BST) are among the most popular and most thoroughly studied data structures for the dictionary problem. There remain however, several outstanding open questions related to the BST model. In particular, what is the best way to adapt a BST in an online fashion, in reaction to a sequence of operations (e.g.\ access, insert, and delete), and what are the theoretical limits of such an adaptation? Does there exist a ``one-size-fits-all'' BST algorithm, asymptotically as efficient as any other dynamic BST algorithm, regardless of the input sequence?

Splay trees have been proposed by Sleator and Tarjan~\cite{ST85} as an efficient BST algorithm, and were shown to be competitive with any \emph{static} BST (besides a number of other attractive properties, such as the \emph{balance}, \emph{working set}, and \emph{static finger} properties). Furthermore, Sleator and Tarjan conjectured splay trees to be competitive with any \emph{dynamic} BST algorithm; this is the famous \emph{dynamic optimality conjecture}~\cite{ST85}. An easier, but similarly unresolved, question asks whether such a dynamically optimal algorithm exists at all. We refer to \cite{iacono_pursuit} for a survey of work related to the conjecture.

A different BST algorithm (later called \greedyfuture) has been proposed independently by Lucas~\cite{Luc88} and by Munro~\cite{Mun00}. \greedyfuture is an offline algorithm: it anticipates future accesses, preparing for them according to a greedy strategy. In a breakthrough result, Demaine, Harmon, Iacono, Kane, and P\v{a}tra\c{s}cu (DHIKP) transformed \greedyfuture into an online algorithm (called here \greedy), and presented a geometric view of BST that facilitates the analysis of access costs (while abstracting away many details of the BST model). 

At present, our understanding of both splay trees and \greedy is incomplete. For splay trees, besides the above-mentioned four properties (essentially subsumed\footnote{Apart from a technicality for working set, that poses no problem in the case of splay trees and \greedy.} by a single statement called the access lemma), a few other corollaries of dynamic optimality have been shown, including the \emph{sequential access}~\cite{tarjan_sequential} and the \emph{dynamic finger}~\cite{finger1, finger2} theorems. The only known proof of the latter result uses very sophisticated arguments, which makes one pessimistic about the possibility of proving even stronger statements.

A further property conjectured for splay trees is a linear cost on deque sequences (stated as the ``deque conjecture'' by Tarjan~\cite{tarjan_sequential} in 1985). Informally, a deque sequence consists of insert and delete operations at \emph{minimum} or \emph{maximum} elements of the current dictionary. Upper bounds for the cost of splay on a sequence of $n$ deque operations are $O(n \alpha(n))$ by Sundar~\cite{sundar} and $O(n \alpha^{\ast}(n))$ by Pettie~\cite{Pettie08}. Here $\alpha$ is the \emph{extremely} slowly growing inverse Ackermann function, and $\alpha^{\ast}$ is its iterated version. A linear bound for splay trees on \emph{output-restricted} deque sequences (i.e.\ where deletes occur only at minima) has been shown by Tarjan~\cite{tarjan_sequential}, and later improved by Elmasry~\cite{Elmasry04}. 

In general, our understanding of \greedy is even more limited. Fox~\cite{Fox11} has shown that \greedy satisfies the access lemma and the sequential access theorem, but no other nontrivial bounds appear to be known. One might optimistically ascribe this to a (relative) lack of trying, rather than to insurmountable technical obstacles. This motivates our attempt at the deque conjecture for \greedy.

As mentioned earlier, a deque sequence consists of insert and delete operations. In the tree-view, e.g.\ for splay trees, such operations have a straightforward implementation. Unfortunately, the geometric view in which \greedy can be most naturally expressed only concerns with accesses. Thus, prior to our work there was no way to formulate the deque conjecture in a managable way for \greedy.

\paragraph{Our contributions.} We augment the geometric model of DHIKP to allow insert and delete operations (exemplified by the extension of the \greedy algorithm), and we show the offline and online equivalence of a sequence of operations in geometric view with the corresponding sequence in tree-view. This extended model allows us to formulate the deque conjecture for \greedy. We transcribe the geometric view of \greedy in matrix form, and we apply the forbidden-submatrix technique to derive the quasilinear bound $O(m2^{\alpha(m,m+n)} +n)$ on the cost of \greedy, while serving a deque sequence of length $m$ on keys from $[n]$.

We also prove an $O(m+n)$ upper bound for the special case of output-restricted deque sequences. We find this proof considerably simpler than the corresponding proofs for splay trees, and we observe that a slight modification of the argument gives a new (and perhaps simpler) proof of the sequential access theorem for \greedy.

As a further application of the insert-delete model we show through a reduction to sorting that for \emph{any} BST algorithm, most representatives from a set $U$ of permutations on $[n]$ have an access cost of $\Omega(\log{|U|} + n)$. In particular, this implies that a random permutation of $[n]$ has access cost $\Omega(n\log{n})$ with high probability. 
A similar result has been shown by Wilber~\cite{wilber} for random access sequences (that might not be permutations). 
Our proof is self-contained, not relying on Wilber's BST lower bound. 
Permutation access sequences are important, since it is known that the existence of a BST algorithm that is constant-competitive on permutations implies the existence of a dynamically optimal algorithm (on arbitrary access sequences).

\paragraph{Related work.} A linear cost for deque sequences is achieved by the multi-splay algorithm~\cite[Thm~3]{Derryberry09propertiesof} in the special case when the initial tree is empty; by contrast, the results in this paper make no assumption on the initial tree.

Most relevant to our work is the deque bound of Pettie for splay trees~\cite{Pettie08}. That result relies on bounds for Davenport-Schinzel sequences, which can be reformulated in the forbidden-submatrix framework. Indeed, the use of forbidden-submatrix theory for proving data structure bounds was pioneered by Pettie, who reproved the sequential access theorem for splay trees~\cite{pettie_DS} (among other data structure results). Our application of forbidden-submatrix theory is somewhat simpler and perhaps more intuitive: the geometric view of \greedy seems particularly suitable for these types of arguments, as the structure of BST accesses is readily available in a matrix form, without the need for an extra ``transcribing'' step. 

\section{\label{sec:GeoFormulation}Geometric Formulation of BST with Insertion/Deletion}

In this section we extend the model of DHIKP \cite{DemaineHIKP09} to allow for insertions and deletions. After defining our geometric model, we prove the equivalence of the arboreal (i.e.\ tree-view) and the geometric views of BSTs.

\subsection{Rotations and Updates}

\begin{figure}
\begin{center} 
\includegraphics[scale=0.1,  trim=0 1.5cm 0 2cm]{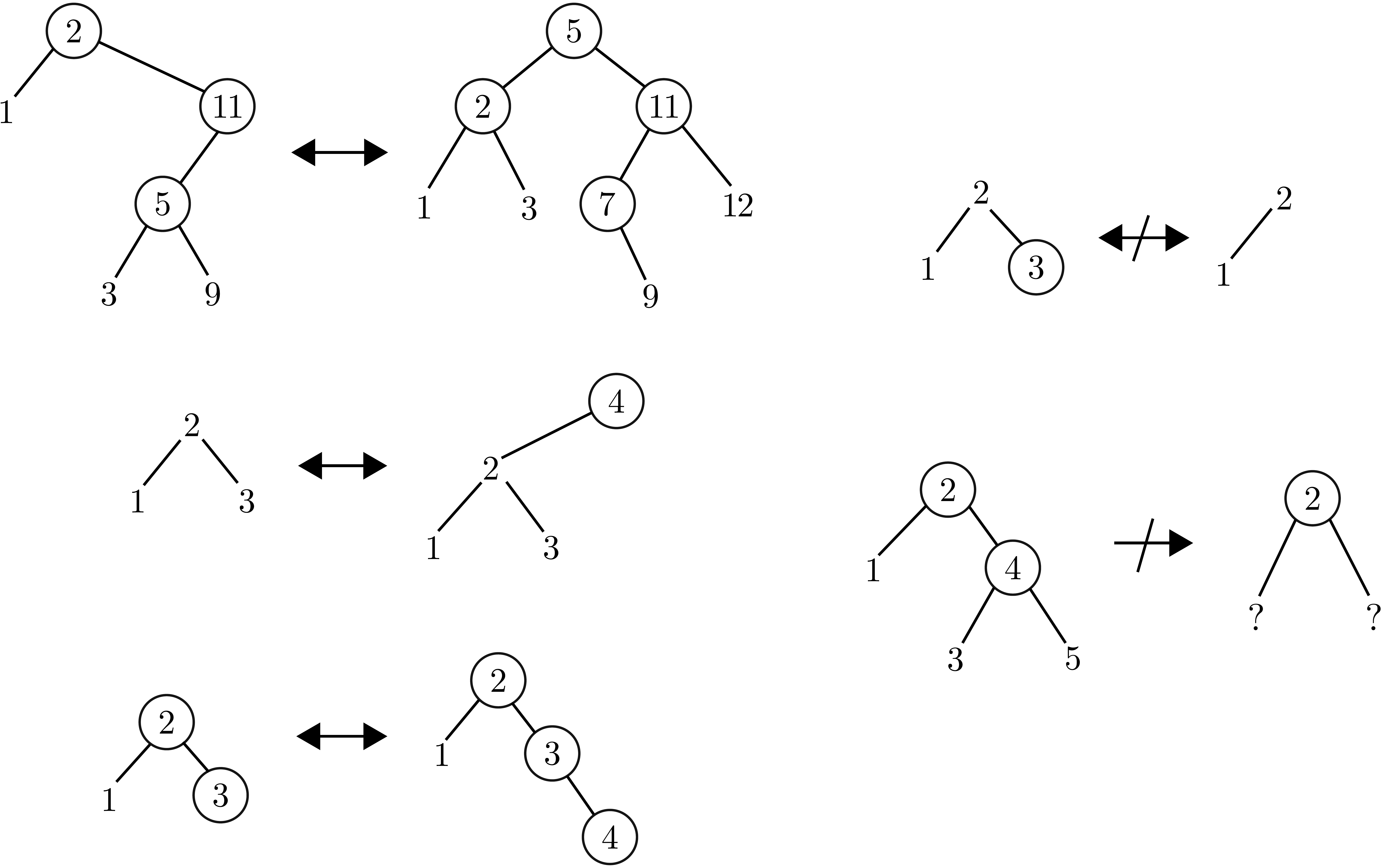}
\end{center}
	\protect\caption{(left) Examples of valid insert/delete operations. Circled elements indicate $\tau$ and $\tau'$; (right) Examples of invalid operations: $\tau$ does not contain root (above) and $\tau'$ cannot link all pendant trees (below).\label{fig1}}
\end{figure}

\begin{definition}
	[Valid Reconfiguration] Given a BST $T_{1}$, a (connected) subtree
	$\tau$ of $T_{1}$ containing the root, and a tree $\tau'$ on the
	same nodes as $\tau$, except that one node may be missing or newly
	added, we say that $T_{1}$ can be reconfigured by an operation $\tau\rightarrow\tau'$
	to another BST $T_{2}$ if $T_{2}$ is identical to $T_{1}$ except
	for $\tau$ being replaced by $\tau'$, meaning that the child
	pointers of elements not in $\tau$ do not change. The cost of the
	reconfiguration is $\max\{|\tau|,|\tau'|\}$.
\end{definition}
This definition differs from \cite[Def.2]{DemaineHIKP09} in that
$\tau'$ need not be defined on the same nodes as $\tau$. Note
that, according to the definition, if an operation $\tau\rightarrow\tau'$
changes a child pointer of an element $x$, then $x\in\tau$. See Figure~\ref{fig1} for examples.

\begin{definition}
	[Execution of Update Sequence]Given an update sequence 
	\[S=\langle(s_{1},\mathtt{op}_{1}),(s_{2},\mathtt{op}_{2}),\dots,(s_{m},\mathtt{op}_{m})\rangle, \text{  where  } \mathtt{op}_{i}\in\{\mathtt{access},\mathtt{insert},\mathtt{delete}\},\]
	we say that a BST algorithm executes $S$ by an execution $E=\langle T_{0},\tau_{1}\rightarrow\tau_{1}',\dots,\tau_{m}\rightarrow\tau_{m}'\rangle$
	if all reconfigurations $\tau_{t}\rightarrow\tau_{t}'$ transforming
	$T_{t-1}$ to $T_{t}$ are valid, and for all $t$
	\begin{itemize}
		\item if $\mathtt{op}_{t}=\mathtt{access},$ then $s_{t}\in\tau_{t}$ and
		$\tau_{t}' = \tau_{t}$ as a set,
		\item if $\mathtt{op}_{t}=\mathtt{insert},$ then $\tau_{t}' = \{s_{t}\} \dot{\cup} \tau_{t}$ as a set,
		\item if $\mathtt{op}_{t}=\mathtt{delete},$ then $\tau_{t} = \{s_{t}\} \dot{\cup} \tau_{t}'$ as a set.
	\end{itemize}

	We also say that $E$ executes $S$. 
	The\emph{ cost of execution of} $E$
	is the sum over all reconfiguration costs. If an element $x\in\tau_{t}\cup\tau_{t}'$,
	we say that $x$ is \emph{touched }at time $t$. 
	
\end{definition}

\vspace{-0.1in}

We assume that we work over the set $[n]$. Each element can be inserted
or deleted many times, but insertions and deletions on the same element
must be alternating. We also assume that every element is accessed
or updated at least once.

\subsection{Valid Sets}
\begin{definition}
	[Geometric View of Update Sequence] The geometric view of an update
	sequence $S$ is a point set $P(S)=A(S)\dot{\cup}I(S)\dot{\cup}D(S)$
	in the integer grid $[n]\times[m]$ consisting of access points $A(S)=\{(s_{t},t)\mid\mathtt{op}_{t}=\mathtt{access}\}$,
	insertion points $I(S)=\{(s_{t},t)\mid\mathtt{op}_{t}=\mathtt{insert}\}$,
	and deletion points $D(S)=\{(s_{t},t)\mid\mathtt{op}_{t}=\mathtt{delete}\}$.
	Update points are $U(S)=I(S)\dot{\cup}D(S)$.
\end{definition}
We usually omit the parameter $S$ and simply write $A,I,D,U$ when the choice of $S$ is clear from context.  
We denote the $x$-coordinate and $t$-coordinate of a point $p$
by $(p_{x},p_{t})$. By element $x$, we mean the column $x$. By
time $t$, we mean the row $t$.

\begin{definition}
	[Valid Point]
	\label{def:valid}
	Given a point set $P(S)$ in the integer grid $[n] \times [m]$, let $p$ be a point ($p$ may not be in $P(S)$), and let $p',p'' \in U(S)$ 
	denote the update points nearest to $p$, below (resp.\ above) $p$, i.e.\ $p'_x = p''_x = p_x$, and $p'_t < p_t < p''_{t}$. 
	One or both of $p'$ and $p''$ might not exist.	We say that $p$ is \emph{valid} in $P(S)$, iff:
	\begin{itemize}
	\item $p \notin U(S)$, $p' \in I(S)$ (or does not exist), and $p'' \in D(S)$ (or does not exist), or
	\item $p \in I(S)$, $p' \in D(S)$ (or does not exist), and $p'' \in D(S)$ (or does not exist), or
	\item $p \in D(S)$, $p' \in I(S)$ (or does not exist), and $p'' \in I(S)$ (or does not exist).
	\end{itemize}
\end{definition}

Let $T_{t}$ denote the resulting tree at time $t$ during an execution of the BST algorithm $E$ on the update sequence $S$. Observe that Definition~\ref{def:valid} allows elements to be accessed or deleted without having been inserted before. Such elements are (implicitly) in the initial tree $T_0$.

\begin{fact}
	\label{fact:valid}A point $x$ can be touched at time $t$ iff $(x,t)$ is valid. 
\end{fact}

\begin{wrapfigure}[19]{l}[0.3\textwidth]{0.7\textwidth}
\begin{center}  
\includegraphics[width=0.33\textwidth, trim=0 1.5cm 0 3.8cm]{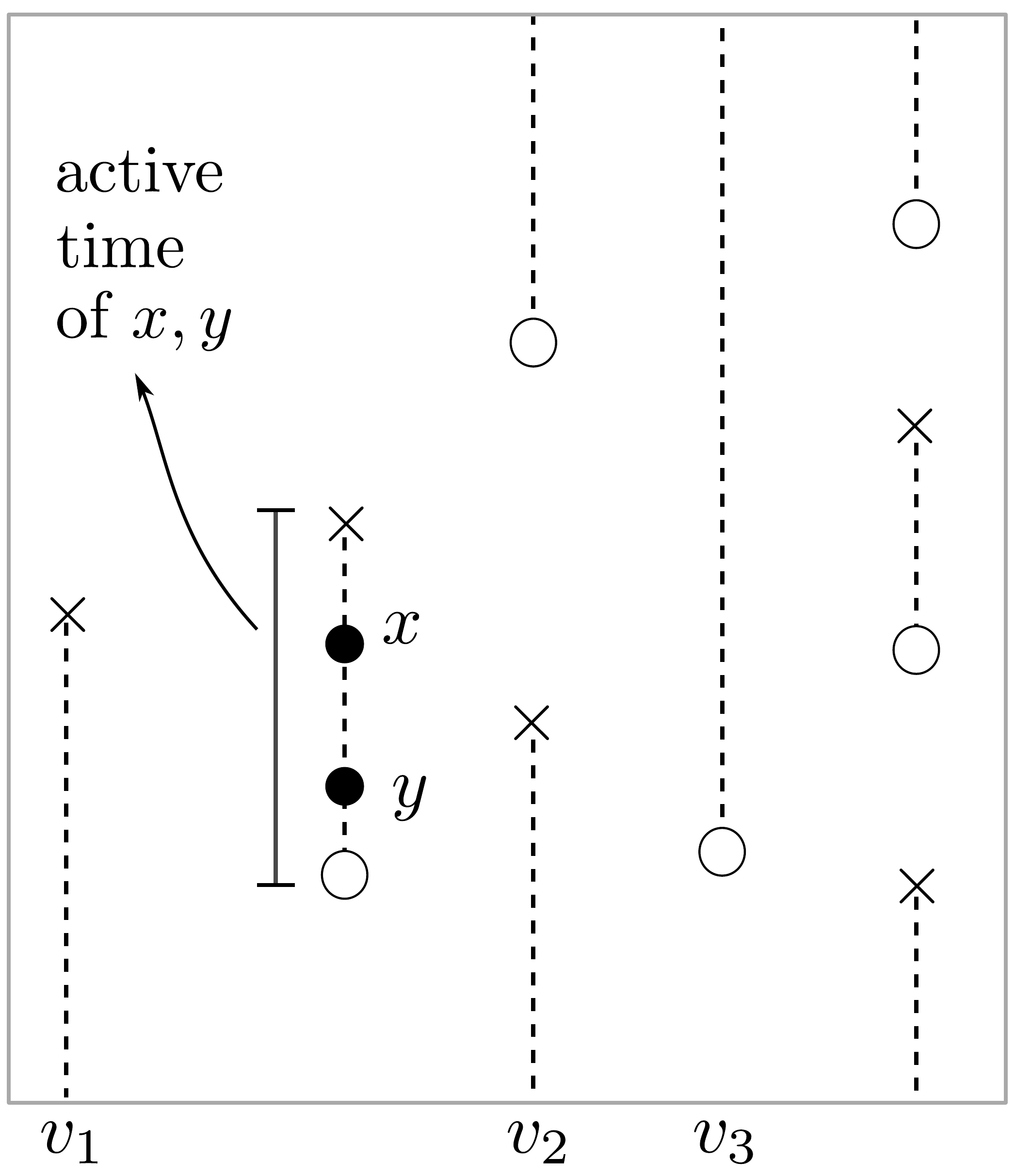}
\end{center}
	\protect\caption{A point set with insert ($\ocircle$) and delete ($\times$) points. Dashed lines indicate valid points. Observe that $\suc(x) = v_3$, $\suc(y) = v_2$, and $\pred(x) = \pred(y) = v_1$.\label{fig2}}
\end{wrapfigure}

Suppose that $(x,t)$ is valid. If $(x,t)$ is a deletion point, then $x$ is in $T_{t-1}$ but not $T_{t}$, and it is touched. If $(x,t)$ is an insertion point, then $x$ is in $T_{t}$ but not $T_{t-1}$, and it is touched. If $(x,t)$ is not an update point, then $x$ is in both trees, and might or might not be touched. See Figure~\ref{fig2} for an illustration.

\begin{definition}
	[Predecessor/Successor of a Point]Given $P(S)$, the predecessor $\pred(p)$
	of a point $p$ is the largest element $x'$ smaller than $p_{x}$
	such that $(x',p_{t})$ is valid. We also write $\pred(p)=(x',p_{t})$ as a point. The successor $\suc(p)$ of $p$
	is symmetrically defined.
\end{definition}

\begin{definition}
	[Valid Set]A point set $P\supseteq P(S)$ is \emph{valid }iff every
	point $p\in P$ is valid.
\end{definition}

For any node $x$ in a tree $T$, let $\pred_{T}(x)$ and $\suc_{T}(x)$ denote the predecessor, respectively successor of $x$
in $T$. The following lemma shows
that points in a valid set, and their predecessor and successor, are
associated with nodes in the tree at the corresponding time.

\begin{lemma}
	\label{valid set: pred/succ}Let $P\supseteq P(S)$ be a valid point
	set, and $E$ executes $S$. For any $p\in U(S)$, we have $\pred(p)=\pred_{T_{p_{t}}}(p_{x})$ and $\suc(p)=\suc_{T_{p_{t}}}(p_{x})$.\end{lemma}
\begin{proof}
	Let $x'=\pred(p)$ and hence $(x',p_{t})$ is valid by definition. By Fact~\ref{fact:valid}, $x'$ can be touched at time $p_t$. Since $x'$ is not an updated element,
	we have $x'\in T_{p_{t}}$. Moreover, $x'$ is
	the closest element on the left of $p_{x}$ at this time. So $x'=\pred_{T_{p_{t}}}(p_{x})$.
	The proof for successor is symmetric.\end{proof}
\begin{definition}
	[Active Time of Points]Let $p$ be a point in a valid point set $P\supseteq P(S)$.
	The active time $act(p)$ of $p$ is the maximal consecutive interval
	of time $[t_{ins}(p),t_{del}(p)]$ containing $p_{t}$ such that,
	for all $t\in act(p)$, $(p_{x},t)$ is valid. We call $t_{ins}(p)$
	insertion time of $p$, and $t_{del}(p)$ deletion time of $p$.
\end{definition}

\subsection{Arboreally Satisfied Set}
\begin{definition}
	[Geometric View of BST Execution]The geometric view of a BST execution
	$E=\langle T_{0},\tau_{1}\rightarrow\tau_{1}',\dots,\tau_{m}\rightarrow\tau_{m}'\rangle$
	of some update sequence $S$ is the point set $P(E)=\{(x,t)\mid x\in\tau_{t}\cup\tau_{t}'\}$
	in the integer grid, indicating which element is touched at which
	time. Note that $P(E)\supseteq P(S)$. 
\end{definition}

\begin{definition} 
	[Arboreally Satisfied Set] \label{def:ASS}A valid point set $P\supseteq P(S)$ is\emph{
		(arboreally) satisfied }iff the following holds:
	\begin{itemize}
		\item For each pair $p,q\in P$ that are both active from time $p_{t}$
		to $q_{t}$ (called an active pair), either 
		both $p$ and $q$ lie in the same vertical/horizontal line, or
		there is a point $r\in\square_{pq} \cap P \setminus \{p,q\}$.
		If $r$ is on the bottommost row of $\square_{pq}$, then $r$ cannot be a deletion point.
		If $r$ is on the topmost row of $\square_{pq}$, then $r$ cannot be an insertion point.
		\item For each update point $p\in U$, if both $\pred(p)$ and $\suc(p)$
		exist, then either $\pred(p)$ or $\suc(p)$ is also
		in $P$. 
	\end{itemize}
\end{definition}
The first condition is almost the same as the one in \cite[Def.\ 2.3]{DemaineHIKP09}
but focused only on \emph{active pairs} (they are active from $p_t$ to $q_t$), and with additional technical condition due to update points.
The second condition says that
if the updated element is not the current minimum/maximum, then one of its
adjacent elements must be touched. 

Note that if there are no update points, then all points are active
the whole time and our definition is equivalent to  \cite[Def.\ 2.3]{DemaineHIKP09}. We defer the proof of the following fact to the appendix.
\begin{fact}
	\label{fact:sat on sides}Suppose that $P$ is satisfied. Then, for each
	pair $p,q\in P$ which are both active from time $p_{t}$ to $q_{t}$ and $p_t<q_t$,
	there exists a point in $P\setminus \{p,q\}$ on a side of $\square_{pq}$ incident to $p$, that is either a non-deletion point, or the corner $(p_x,q_t)$.
	Similarly, there exists a point in $P\setminus \{p,q\}$ on a side of $\square_{pq}$ incident to $q$, that is either a non-insertion point, or the corner $(q_x,p_t)$.
\end{fact}

\section{Equivalence of Arboreal and Geometric Views}
\label{sec:equiv}
In this section we prove the following theorem:
\begin{theorem}
	\label{thm:geo equiv tree}A point set $P$ is satisfied iff $P=P(E)$
	for some BST execution $E$.
\end{theorem}

The first direction of the proof involves considering a BST algorithm and showing that it generates a satisfied point set (tree to geometry). The second direction is showing how to convert a satisfied point set to a BST algorithm (geometry to tree).

\subsection{Tree to Geometry}
\label{sec:ttog}
\begin{lemma}
	\label{consec ancestor}Let $x$ and $z$ be elements with consecutive
	values in a BST $T$, with $x<z$. Then one of $x$ and $z$ is an ancestor of
	the other.\end{lemma}
\begin{proof}
	Suppose not. Then the lowest common ancestor of $x$ and $z$ is another
	element $y$. We know $x<y<z$ which is a contradiction.\end{proof}
\begin{lemma}
	\label{update -> touch pred/succ}Suppose that $y$ is not the minimum
	or maximum element in a BST $T$. To insert or delete $y$ in $T$,
	either $\pred_{T}(y)$ or $\suc_{T}(y)$ must be touched.\end{lemma}

\begin{lemma}
	\label{lem:tree to geo}For any execution $E$, a point set $P(E)$
	is satisfied.\end{lemma}
\begin{proof}
	There are two conditions that need to be checked. 
	
	For the first condition, let $p,q$ be a pair of points in $P(E)$
	active from time $p_{t}$ to $q_{t}$. Suppose that $p,q$ violate
	the condition. Hence, they are not vertically or horizontally aligned.
	We assume that $p_{t}<q_{t}$ and $p_{x}<q_{x}$. Since $p_{x}$ and
	$q_{x}$ are active at time $p_{t}$, by Fact~\ref{fact:valid} and
	the statement below the fact, they exist in the tree $T_{p_{t}}$.
	Hence, a lowest common ancestor $a$ of $p_{x}$ and $q_{x}$ in $T_{p_{t}}$
	is well-defined. There are two cases. 
	
	If $a=p_{x}$, then $p_{x}$ is an ancestor of $q_{x}$. Since $\square_{pq}$
	is not satisfied, $q_{x}$ is not touched from time $p_{t}$ to $q_{t}-1$
	and $p_{x}$ remains an ancestor of $q_{x}$ right before time $q_{t}$.
	Thus, to touch $q_{x}$ at time $q_{t}$, $p_{x}$ must be touched,
	and so $(p_{x},q_{t})\in\square_{pq}$. Only insertion point can be
	in the topmost row of unsatisfied $\square_{pq}$. So $(p_{x},q_{t})$
	an insertion point. But this implies that $p$ and $q$ are not active
	pair, which is a contradiction.
	
	If $a\neq p_{x}$, then $a$ must be touched at time $p_{t}$. As
	$a$ has value between $p_{x}$ and $q_{x}$, we have $(a,p_{t})\in\square_{pq}$.
	Since $\square_{pq}$ is not satisfied, $(a,p_{t})$ is a deletion
	point and, moreover, $p_{x}$ must be its predecessor. Hence $p_{x}$
	becomes an ancestor of $q_{x}$ right after time $p_t$ and we can use the previous argument
	again. 
	
	For the second condition, suppose that $p\in U$ is an update point.
	That is, we update $p_{x}$ in the BST $T_{p_{t}}$. If both $\pred(p)$
	and $\suc(p)$ exist, then $p_{x}$ is not a minimum or maximum in
	$T_{p_{t}}$. By Lemma \ref{update -> touch pred/succ}, either $\pred_{T_{p_{t}}}(p_{x})$
	or $\suc_{T_{p_{t}}}(p_{x})$ is touched at time $p_{t}$. By Lemma
	\ref{valid set: pred/succ}, $\pred_{T_{p_{t}}}(p_{x})=\pred(p)$ and
	$\suc_{T_{p_{t}}}(p_{x})=\suc(p)$, and we are done.
\end{proof}

\subsection{Geometry to Tree}
\label{sec:gtoffline}
Now we show how to convert a valid point set to an offline algorithm first. We need the following lemma, which is essentially a converse of Lemma
\ref{update -> touch pred/succ}, saying that if we touch either $\pred_{T}(y)$
or $\suc_{T}(y)$, then we can insert or delete $y$. We defer the proofs of the following two statements to the appendix.
\begin{lemma}
	\label{touch pred/succ -> can update}Suppose either $\pred_{T}(y)$
	or $\suc_{T}(y)$ is in a subtree $\tau$ containing the root of $T$, or $y$ is the minimum or maximum element in $T$. 
	Then (i) any reconfiguration $\tau\rightarrow\tau'$, where 
	$\tau'=\tau \dot{\cup} \{y\}$
	as a set, is valid, 
	and (ii) any reconfiguration
	$\tau\rightarrow\tau'$, where $\tau=\tau' \dot{\cup} \{y\}$ as a set,
	is valid. 
	\end{lemma}

\begin{lemma}
	[Offline Equivalence]\label{thm:offline geo to tree}For any satisfied
	set $X$, there is a point set $P(E)=X$ for some execution $E$.
	We call $E$ a tree view of $X$.\end{lemma}
By Lemma \ref{lem:tree to geo} and \ref{thm:offline geo to tree},
this concludes the proof of Theorem \ref{thm:geo equiv tree}.

Observe that if $X = P(E)$, the quantity $|X|$ is exactly the execution cost of $E$.

\subsection{Geometry to Tree: Online}

The discussion in \S\,\ref{sec:gtoffline} assumes that a satisfied set $X$ is available all at once, and we show that there exists an execution $E$ (i.e.\ an offline BST algorithm) whose point set $P(E)$ is exactly $X$. 

We call an \emph{online geometric algorithm} an algorithm that, given a geometric update sequence $P(S) \subseteq [n] \times [m]$, outputs a satisfied superset $P \supseteq P(S)$, with the condition that both the input and output are revealed \emph{row-by-row} (i.e.\ the decision on which points to touch can depend only on the current and preceding rows of the input). We remark that \greedy (as extended in \S\,\ref{sec:greedy}) is such an algorithm.

Analogously, by an online BST algorithm we mean a procedure that, given an initial set $S_0 \subseteq [n]$, and an update sequence $S$, outputs an execution $E$, with the condition that both the input and output are revealed \emph{item-by-item} (i.e.\ the decision on which reconfiguration to perform can depend only on the current and preceding update operations).

\begin{theorem}
[Online Equivalence]\label{thm: online}For any online geometric algorithm $\mathcal{A}$, there exists
	an online BST algorithm $\mathcal{A'}$ such that, on any update sequence, the
	cost of $\mathcal{A'}$ is bounded by a constant times the cost of $\mathcal{A}$.
\end{theorem}

The proof of Theorem~\ref{thm: online} is an adaptation of the proof of Lemma $2.3$ in \cite{DemaineHIKP09} to the new geometric setting, and is analogous to the proof of Lemma~\ref{thm:offline geo to tree}. We omit the proof in this extended abstract. 

\section{Defining \greedy with Insertion/Deletion}
\label{sec:greedy}
\greedy is an online algorithm for constructing a satisfied set
given an update sequence $S$. At each time $t$, \greedy minimally
satisfies the point set up to time $t$. Having defined satisfied
sets when there are update points, we naturally obtain the extension of \greedy that
can handle insertions and deletions.

\begin{wrapfigure}[18]{l}[0.3\textwidth]{0.68\textwidth}
\begin{center}  
\includegraphics[width=0.35\textwidth, trim=0 1.5cm 0 4cm]{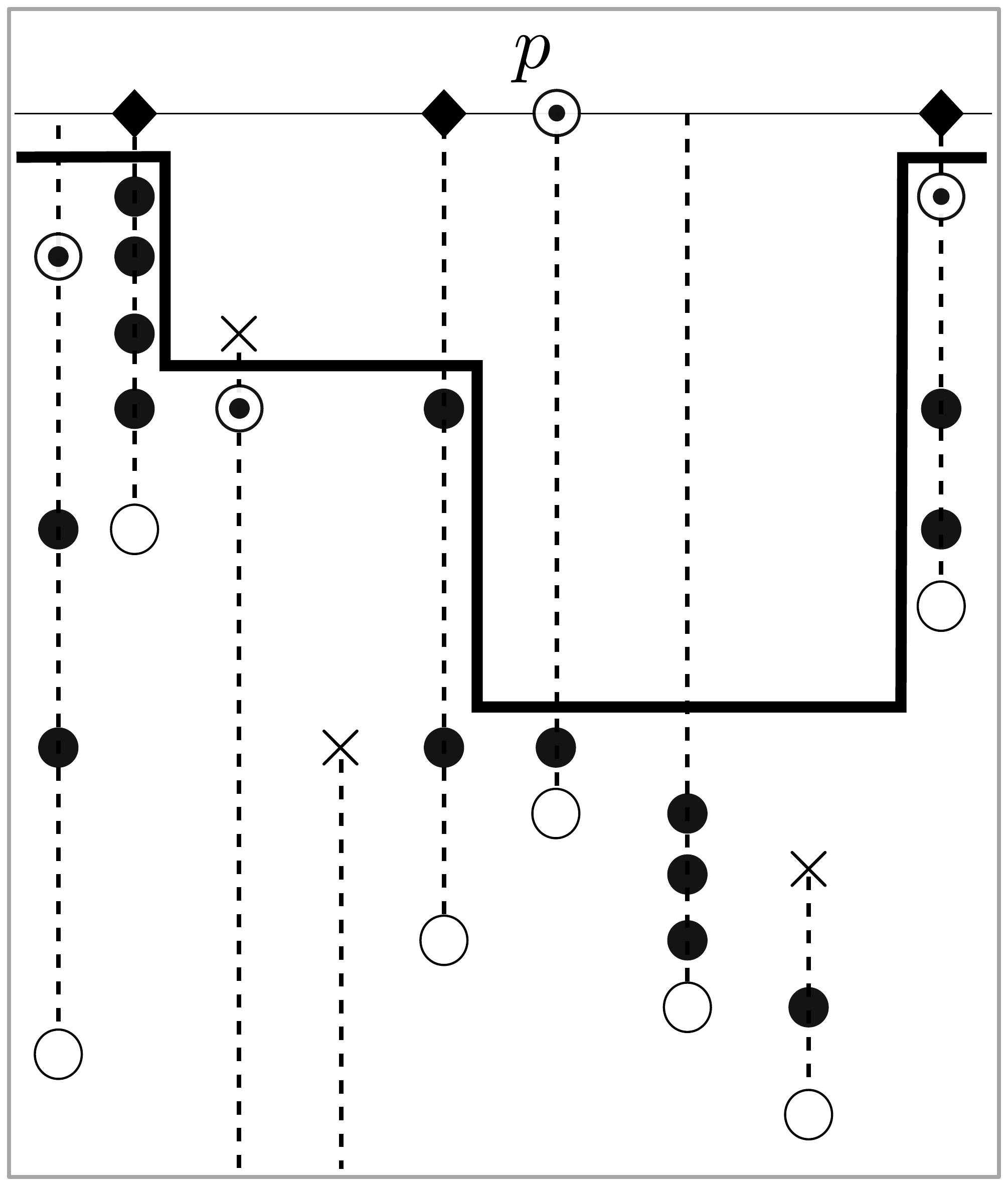}
\end{center}
	\protect\caption{A \greedy execution with insert ($\ocircle$), delete ($\times$), access ($\odot$), touched ($\newmoon$) points, and touched points at time $p_t$ ($\Diamondblack$). Thick line shows stair of $p$. Observe that a non-(min/max) insert or delete must access a neighbor as well.\label{fig5}}
\end{wrapfigure}

We develop some notation for describing the algorithm. A rectangle $\square_{pq}$
is \emph{unsatisfied} if there is no other point in the proper (closed) rectangle
formed by points $p$ and $q$. We say that $p$ and $q$ are an \emph{active
	pair }if they are active from time $p_{t}$ to $q_{t}$. The \emph{stair}
of point $p$ is denoted by $\stair(p)=\{p\}\cup\{q\mid\square_{pq}$
is unsatisfied rectangle formed by an active pair $p$ and $q$ where
$q$ is below $p\}$. The \emph{stair} of element $x$ at time $t$
is the stair of the point $(x,t)$. \emph{Satisfying/touching} $\stair(x,t)$
means visiting/touching, at time $t$, the elements of points in the stair:
$\{(q_{x},t)\mid q\in \stair(x,t)\}$. These elements are then added to the row at time $t$.

\begin{fact}
	Touching the stair $\stair(p)$ is to minimally satisfy the point $p$. 
\end{fact}
Therefore, when \greedy gets an access point $p$, it touches only $\stair(p)$. 
For an update point $p$, if $p$ is not the minimum
or maximum, then \greedy chooses the smaller set between $\stair(p)\cup \stair(\pred(p))$
and $\stair(p)\cup \stair(\suc(p))$. This is because of the second
condition of satisfied set. If $p$ is the minimum or maximum, then
\greedy just touches $\stair(p)$. The execution of \greedy is illustrated in Figure~\ref{fig5}.


The following observation is useful for deque sequences.
For insertion point $p$, observe that $\stair(p)=\{p\}$ because the
active time of $p$ begins at time $p_{t}$ itself (for any
point $q$ below $p$, $p$ and $q$ are not an active pair by definition). 
\begin{fact}
	To insert $p$ such that $p$ is the minimum or maximum, \greedy touches only
	$p$.
\end{fact}

\section{Performance of \greedy on Deque Sequences}
\label{sec:grdeq}
\begin{definition}
	[Deque Sequence]An update sequence is a \emph{deque sequence} if
	it has only insertions and deletions at the current minimum or maximum
	element, and no access operations. 
\end{definition}

\begin{definition}
	[Output-restricted]A deque sequence is \emph{output-restricted} if
	it has deletions only at minimum elements.
\end{definition}

\begin{wrapfigure}[18]{l}[0.2\textwidth]{0.5\textwidth}
\begin{center}  
\includegraphics[width=0.28\textwidth, trim=0 1.5cm 0 3cm]{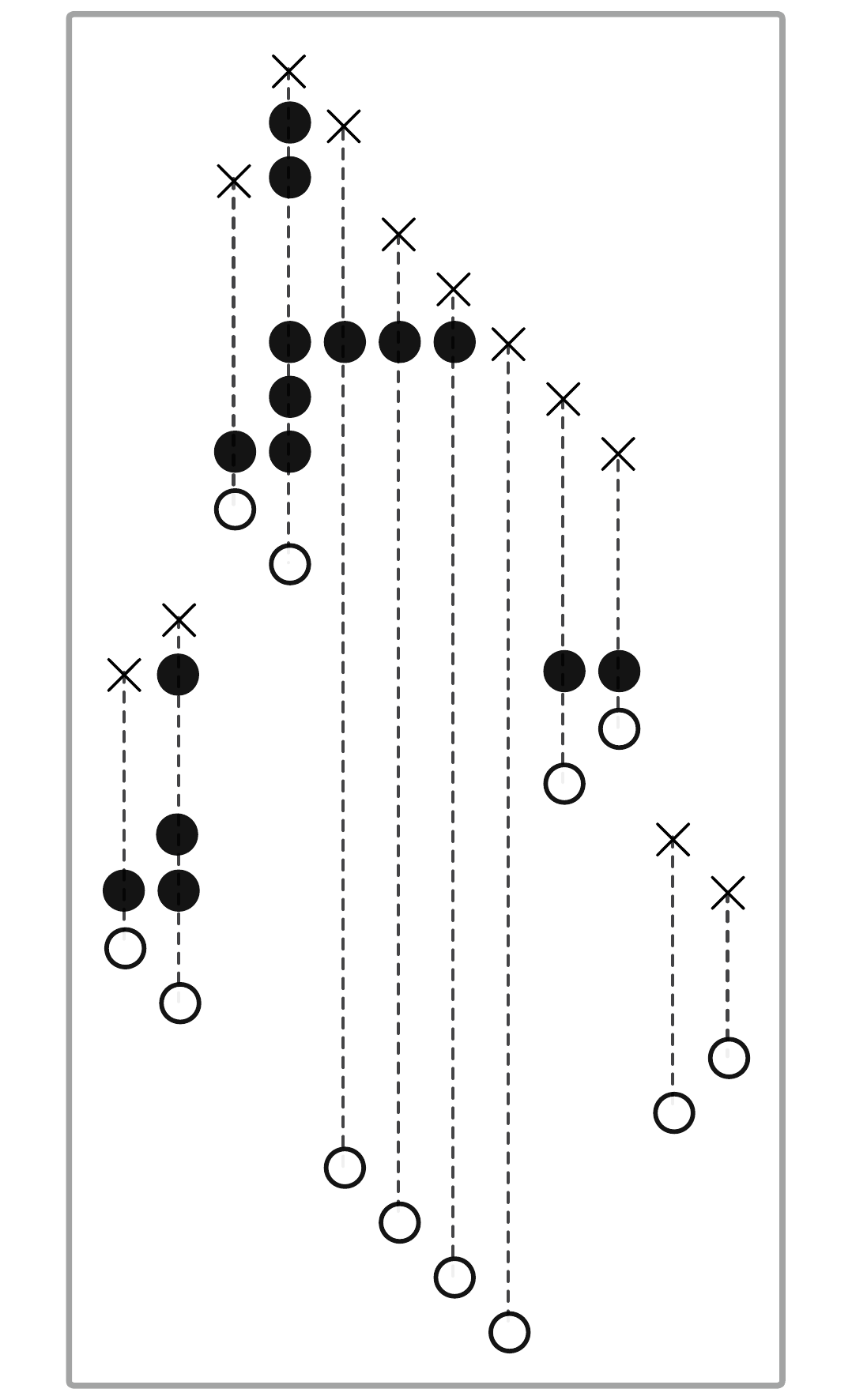}
\end{center}
	\protect\caption{Sample execution of \greedy on a concentrated deque sequence with insert ($\ocircle$), delete ($\times$), and touched ($\newmoon$) points. Dashed lines show the active times of elements.\label{fig3}}
\end{wrapfigure}

\begin{theorem}\label{greedy_deque}
The cost of executing a deque sequence on $[n]$ of length $m$ by \greedy
is at most $O(m2^{\alpha(m,n+m)}+n)$, where $\alpha$ is the inverse Ackermann function.
\end{theorem}

\begin{theorem}\label{greedy_linear}
The cost of executing an output-restricted deque sequence on $[n]$ of length $m$ by \greedy
is at most $24m+12n$.
\end{theorem}

\paragraph{Remark.} The bound in Theorem~\ref{greedy_linear} refers to the cost of the online \emph{geometric} \greedy. In the online tree-view equivalent the constants can be larger, hinging on the details of Theorem~\ref{thm: online}, but the bound remains of the form $O(m+n)$.

The rest of this section is devoted to the proofs of Theorems~\ref{greedy_deque} and \ref{greedy_linear}.

\subsection{Concentrated Deque Sequences}

We first reduce the analysis of \greedy on any deque sequence to that on a special type of deque sequence that we call a concentrated deque sequence. Recall that in a deque sequence we can delete only the current minimum or
maximum. We define two sets of elements as follows: let $L_{t}$
be the set of elements which are deleted (from the left) before time $t$ when they
were the minimum at their deletion time, and $R_{t}$ be the
set of elements which are deleted (from the right) before time $t$ when they were \emph{not the minimum} at their deletion time. Observe that $L_t \cap R_t = \varnothing$.
\begin{definition}
	[Concentrated Deque Sequence]A deque sequence is concentrated if,
	for any time $t$, if the inserted element $x$ is the minimum, then
	$y<x$ for all $y\in L_{t}$, and if $x$ is the maximum, then $x<y$
	for all $y\in R_{t}$.\end{definition}

Note that the definition implies that each element in a concentrated deque sequence can be inserted and deleted at most once. We defer the proof of the following lemma to the appendix.

\begin{lemma}
	\label{lem:reduction concentrated}For any deque sequence $S$, there
	is a concentrated deque sequence $S'$ such that the execution of
	any BST algorithm on $S'$ and $S$ have the same cost.\end{lemma}

\subsection{\greedy on a Concentrated Deque Sequence}


Now we analyze the performance of \greedy on concentrated deque sequences (see Figure~\ref{fig3} for an example). Because of Lemma~\ref{lem:reduction concentrated}, we can view the points touched by \greedy as an $(m \times (n+m))$ binary matrix (i.e.\ with entries $0$ and $1$), with all touched points represented as ones, and all other grid elements as zeroes. Notice that the number of columns is $n+m$ instead of $n$ because of the reduction in Lemma~\ref{lem:reduction concentrated} which allows each element to be inserted and deleted at most once. We further observe that if a deque sequence is output-restricted, then the transformation of Lemma~\ref{lem:reduction concentrated} yields a concentrated deque sequences that is similarly output-restricted.

\begin{definition}[Forbidden Pattern]
A binary matrix $M$ is said to \emph{avoid} a binary matrix $P$ (called a pattern) if there exists no submatrix $M'$ of $M$ with same dimensions as $P$, such that for all $1$-entries of $P$, the corresponding entry in $M'$ is $1$ (the $0$-entries of $P$ are ``don't care'' values).
\end{definition}

We denote by $\mathrm{Ex}(P,m,n)$ the largest number of $1$s in an $(m \times n)$ matrix $M$ that avoids pattern $P$. In this work, we refer to the following patterns (as customary, we write dots for $1$-entries and empty spaces for $0$-entries).

\begin{align*}
P_5 = \left( \begin{matrix}
  \bullet & \  & \bullet & \  & \bullet \\
  \  & \bullet & \  & \bullet & \ 
 \end{matrix}\right) \quad and \quad
 P_4 = \left( \begin{matrix}
  \bullet & \ & \ & \ \\
  \ & \ & \bullet & \ \\
  \ & \bullet & \ & \ \\\
  \ & \ & \ & \bullet\\
 \end{matrix}\right)
\end{align*} 


	

\begin{lemma}
	\label{lem:deque forbid}The execution of \greedy
	on concentrated deque sequences avoids the pattern $P_{5}$.\end{lemma}
\begin{proof}
	Suppose that $P_5$ appears in the \greedy execution, and name the touched points matched to the $1$-entries in $P_5$ from left to right as
	$a,b,c,d,$ and $e$. 
	
	Let $t > b_t$ be smallest such that $(c_x,t)$ is touched. Then $t \le c_t$ and either $b$ or $d$ must have been deleted within the time interval $[b_t,t]$. Otherwise, any update point in the interval $[b_t,t]$ is outside
	the interval $[b_{x},d_{x}]$ and $c_{x}$ is ``hidden'' by $b$ and $d$ (it cannot be on the stair of any update point). 
	
	Assume w.l.o.g.\ that $b$ is deleted. If $b$ is deleted by a minimum-delete, then $a$ cannot be touched. If $b$ is deleted by a maximum-delete, then $e$ cannot be touched.
	This is because the sequence is concentrated.\end{proof}

\begin{lemma}
	\label{lem:deque linear forbid}The execution of \greedy
	on concentrated output-restricted deque sequences avoids the pattern $P_4$.\end{lemma}
\begin{proof}
	Suppose that $P_4$ appears in the \greedy execution, and name the touched points matched to the $1$-entries in $P_4$ from left to right as
	$a,b,c,$ and $d$. 
	We claim that in order to touch $c$, there has to be a deletion point in the interval $[b_x, d_x]$ in the time interval $[d_t,c_t]$. Otherwise, any deletion point in the time interval $[d_t, c_t]$ is left of $b_x$ (as deletes happen only at the minimum). Furthermore, all insertion points in the time interval $[b_t, c_t]$ must be outside of $[b_x, d_x]$ (since both $b$ and $d$ are active at time $b_t$). We remind that insertion touches nothing else besides the insertion point itself. This means that $c$ cannot be touched: it is ``hidden'' to deletion points on the left of $b_x$ by $b$.
	
	Denote the deletion point in the rectangle $[b_x, d_x] \times [d_t,c_t]$ as $d'$.
	Observe that $a$ is to the left of and above $d'$, and since we only delete minima, $a$ is not active at time $d'_t$. In order to be touched, $a$ must become active after $d'_t$ via an insertion, contradicting that the sequence is concentrated.
  \end{proof}

\begin{fact}
\label{fact:bound P_5} \normalfont\textbf{(\! \cite[Thm $3.4$]{pettie2011generalized} ).}\ \ $\mathrm{Ex}(P_5,u,v) = O(u2^{\alpha(u,v)} +v)$.
	\end{fact}
	
	\begin{fact}
		\label{fact:bound P_4} \normalfont\textbf{(\! \cite[Thm $1.5(5)$]{pettie2011generalized} ).}\ \ $\mathrm{Ex}(P_4,u,v) < 12(u+v)$.
	\end{fact}
	
\paragraph{Proof of Theorem~\ref{greedy_deque}:} By Lemma~\ref{lem:reduction concentrated}, it is enough to analyze
 the cost of \greedy on concentrated deque sequences. This cost is bounded by $O(m2^{\alpha(m,m+n)} +n)$ using Lemma~\ref{lem:deque forbid}
	and Fact~\ref{fact:bound P_5}.

\paragraph{Proof of Theorem~\ref{greedy_linear}:} By Lemma~\ref{lem:reduction concentrated}, it is enough to analyze
 the cost of \greedy on concentrated deque sequences. This cost is bounded by $24m+12n$ using Lemma~\ref{lem:deque linear forbid} and Fact~\ref{fact:bound P_4}.

\paragraph{Remark.} The proof of Theorem~\ref{greedy_linear} can be minimally adjusted to prove the sequential access theorem for \greedy. Sequential access can be simulated as a sequence of minimum-deletions. In this way we undercount the cost by exactly one touched point above each access, which adds a linear term to the bound. 

\section{A Lower Bound on Accessing a Set of Permutations}

Let $U$ be a set of permutations on $[n]$. In this section we prove the following theorem:

\begin{theorem}\label{lowerbound}
Fix a BST algorithm $\mathcal{A}$ and a constant $\epsilon<1$. There exists $U^{'} \subseteq U$ of size $\vert U^{'} \vert \geq (1-\frac{1}{{\vert U \vert}^{\epsilon}})\vert U \vert$ such that $\mathcal{A}$ requires $\Omega(\log \vert U \vert + n)$ access cost on any permutation in $U^{'}$.
\end{theorem}

\begin{proof}
The proof utilizes the  geometric view of insertions, and uses two reductions. We first claim that there exists an algorithm $\mathcal{B}$ that is capable of insertions such that the cost of $\mathcal{A}$ to access a permutation $\pi$ is no less than the cost of $\mathcal{B}$ to insert $\pi$. Note that since $\mathcal{A}$ is accessing $\pi$, all the points are active by definition. We will describe $\mathcal{B}$ in the geometric view simply by requiring that upon inserting $\pi(t)$ at time $t$, $\mathcal{B}$ touches all the points that $\mathcal{A}$ touches while accessing $\pi(t)$ at time $t$. Note that $\mathcal{A}$ touches at least all the points in $\stair(\pi(t),t)$, and $\mathcal{B}$ is required only to touch either $\pred(\pi(t))$ and its stair, or $\suc(\pi(t))$ and its stair (Definition $11$). Since $\pred(\pi(t))$ belongs to $\stair(\pi(t),t)$, one easily sees that $\stair(\pred(p)) \subset \stair(\pi(t),t)$, and this defines a valid insertion algorithm.

We now reduce $\mathcal{B}$ to an algorithm for sorting $\pi$. Just by a traversal of the tree maintained by $\mathcal{B}$ at time $n$, we can produce the sorted order of $\pi$ after incurring a cost of $O(n)$. However, we know that to sort a set $U$ of permutations, any (comparison-based) sorting algorithm must require $\Omega(\log \vert U \vert +n)$ comparisons on at least a $1-\frac{1}{{\vert U \vert}^{\epsilon}}$ fraction of the permutations in $U$. To see this, note that the decision tree of any sorting algorithm must have at least $\vert U \vert$ leaves (note that here we are assuming the weaker hypothesis that $\mathcal{A}$ and hence the sorting algorithm, are only designed to work on $U$; they may fail outside $U$). The number of leaves at height at most $(1-\epsilon) \log \vert U \vert$ is at most ${\vert U \vert}^{1-\epsilon}$, and hence at least a $1-\frac{1}{{\vert U \vert}^{\epsilon}}$ fraction require at least $(1-\epsilon) \log \vert U \vert = \Omega( \log \vert U \vert )$ comparisons. Adding the trivial bound of $\Omega(n)$ to scan the input permutation gives us the desired bound.

\paragraph{Remark.} Upper bounds proved for our model do not directly translate into bounds for algorithms. For example, when a new maximum is inserted, this can be done at a cost of one by making the element the root of the tree, respectively, only touching the element inserted. Note that this requires the promise that the element inserted is actually a new maximum. A slight extension makes the model algorithmic. This is best described in tree-view. We put all nodes of the tree in in-order into a doubly-linked list. Then, in the case of an insertion one can actually stop the search once the predecessor or the successor of the new element has been reached in the search because by also comparing the new element with the neighboring list element, one can verify that a node contains the predecessor or successor. Thus at the cost of a constant factor, bounds proved for our model are algorithmic.
\end{proof}

\paragraph{Acknowledgement.} We thank an anonymous reviewer for valuable comments.

\bibliographystyle{plain}
\bibliography{ref}

\newpage 
\appendix

\section{Proof Omitted from Section~\ref{sec:GeoFormulation}}

\subsection{Proof of Fact~\ref{fact:sat on sides}}

	We give the proof only for $p$,	as it is symmetric for $q$.
	
	Since $P$ is satisfied, there is a point $r$ in $\square_{pq}\cap P\setminus\{p,q\}$
	that satisfies $\square_{pq}$. If $r$ is on the horizontal side incident to $p$, then it is not a deletion point, otherwise it would not satisfy $\square_{pq}$
	by the first condition of Definition \ref{def:ASS}. If $r$ is on the vertical side incident to $p$, and not $(p_x,q_t)$, then it is not a deletion point, otherwise $p$ and $q$ would not be an active pair.
	Thus, if $r$ is on a side of $\square_{pq}$ incident to $p$, we are done. Suppose this is not the case, and let $r$ be a point in $\square_{pq}\cap P\setminus\{p,q\}$ satisfying $\square_{pq}$
	such that $t_{ins}(r)$ is minimum. We claim that $t_{ins}(r)\le p_{t}$. 
	
	First, we have $t_{ins}(r)<q_{t}$ because otherwise $t_{ins}(r)=q_{t}$
	and hence $r$ is an insertion point on the topmost row of $\square_{pq}$, and cannot satisfy $\square_{pq}$ due to the first condition of Definition
	\ref{def:ASS}. Then there must be some other point in $\square_{pq}\cap P\setminus\{p,q\}$, whose insertion time is before $q_t$, a contradiction.
	
	Next, suppose $t_{ins}(r)>p_{t}$ then the point $r'=(r_{x},t_{ins}(r))$
	is an insertion point, and, by the second condition of Definition
	\ref{def:ASS}, there is another point $r''\in\square_{pq}$ which
	is either $\pred(r')$ or $\suc(r')$. Note
	that $r''\neq p,q$ because $p_{t}<t_{ins}(r)<q_{t}$. Observe that $t_{ins}(r'') < t_{ins}(r)$, contradicting the choice of $r$. 
	
	Now, since $t_{ins}(r)\le p_{t}$, $p$ and $r$ are both active from
	time $p_{t}$ to $r_{t}$, and we can repeat the same argument as
	long as $r$ is not on the sides of $\square_{pq}$ incident to $p$.

\section{Proof Omitted from Section~\ref{sec:equiv}}

\subsection{Proof of Lemma~\ref{update -> touch pred/succ}}

	Since $y$ is not the minimum or maximum, both predecessor $\pred_{T}(y)$
	and successor $\suc_{T}(y)$ of $y$ exist. Let $x=\pred_{T}(y)$ and
	$z=\suc_{T}(y)$. We consider two cases: insertion and deletion.
	
	For insertion of $y$, before we insert, $y\notin T$ holds. Then $x$ and
	$z$ are consecutive, and, by Lemma \ref{consec ancestor}, we assume
	by symmetry that $x$ is an ancestor of $z$. In particular, $x$
	has right child. After insertion, $x$ and $y$ are consecutive. So,
	by Lemma \ref{consec ancestor}, either $x$ or $y$ is an ancestor
	of the other. If $x$ is an ancestor of $y$, then $x$ must have
	been touched so that we can change pointers below $x$. Otherwise,
	$y$ is an ancestor of $x$ and $x$ cannot have a right child. Thus,
	$x$'s right child pointer must have been changed to null, meaning
	that $x$ has been touched.
	
	For deletion of $y$, suppose that we do not touch both $x$ and $z$.
	If $y$ has at most one child, then one of $x$ and $z$ is an ancestor
	of $y$, but we must touch all the ancestors of $y$, a contradiction.
	If $y$ has two children, then $x$ has no right child and $z$ has
	no left child. Suppose we delete $y$ without touching $x$ and $z$.
	Therefore, $x$ still has no right child and $z$ still has no left
	child even after $y\notin T$. This contradicts Lemma \ref{consec ancestor}.

\subsection{Proof of Lemma~\ref{touch pred/succ -> can update}}	
	
		Suppose that $x=\pred_{T}(y)\in\tau$. The proof when $\suc_{T}(y)\in\tau$
	is symmetric. There are two statements to be proved regarding insertion
	and deletion of $y$, respectively. Let $\child_{R}(x)$ denote a right
	child pointer of element $x$. 
	
	For insertion, we show that $\tau\rightarrow\tau'$ where $\tau'=\tau \dot{\cup}\{y\}$
	is valid. First, we insert $y$ into $\tau$ as a right child of $x$.
	The only pointer changes are: $\child_{R}(y)\gets \child_{R}(x)$ and
	$\child_{R}(x)\gets y$. Finally, we rotate the resulting subtree,
	which includes $y$, to get $\tau'$.
	
	For deletion, we show that $\tau\rightarrow\tau'$ where $\tau=\tau' \dot{\cup}\{y\}$
	is valid. Again, we rotate $\tau$ such that $y$ is a right child
	of $x$, and then remove $y$. The only pointer change is: $\child_{R}(x)\gets \child_{R}(y)$.
	Then we rotate the resulting subtree, which excludes $y$, to get
	$\tau'$.

\subsection{Proof of Lemma~\ref{thm:offline geo to tree}}

	We use the almost same argument as in \cite[Lemma 2.2]{DemaineHIKP09}
	but we need to make sure that we can also update elements while touching
	all points in $X$ exactly. The argument is as follows.
	
	Define the \emph{next touch time} $N(x,t_{0})$ of $x$ at time $t_{0}$
	in $X$ to be the minimum $t$-coordinate of any point in $X$ on
	the ray from $(x,t_{0})$ to $(x,\infty)$. If there is no such point,
	then $N(x,t_{0})=\infty$. 
	
	Let $T_{t}$ be the treap defined on all points $(x,N(x,t))$ active right after time $t$. 
	Recall
	that a treap is a BST on the first coordinate and a heap on the second.
	Let $X_{t}$ denote the set of elements in the row $t$ of $X$. Since
	$T_{t}$ is a treap with heap priority $N(\cdot,t)$, $\tau_{t}=X_{t}\cap T_{t}$
	is connected subtree containing the root of $T_{t}$. So we have $\tau_{t}=X_{t}$,
	and $\tau_{t}=X_{t}\setminus\{y\}$ if we insert $y$ at time $t$,
	as desired.
	
	If there is an update element $y$ in $X_{t}$, and $\pred_T(y)$ and $\suc_T(y)$ exist, by Lemma \ref{touch pred/succ -> can update},
	we just need to show that either $\pred_{T_{t}}(y)\in X_{t}$ or $\suc_{T_{t}}(y)\in X_{t}$.
	Since $X$ is satisfied, either $\pred(y,t)\in X_{t}$ or $\suc(y,t)\in X_{t}$,
	say $\pred(y,t)\in X_{t}$. By Fact \ref{valid set: pred/succ}, $\pred_{T_{t}}(y)=\pred(y,t)$.
	Therefore, $\tau_{t}\rightarrow\tau_{t}'$ is a valid reconfiguration
	where $\tau_{t}\cup\tau_{t}'=X_{t}$.
	
	After, we update $y$ in $\tau_{t}$ and get $\tau_{t}'$, we want
	to get $T_{t+1}$ which is a treap defined on $N(\cdot,t+1)$. To
	get this, we just heapify $\tau_{t}'$ based on $N(\cdot,t+1)$. We
	claim that the whole tree is now $T_{t+1}$. The following argument
	is exactly same as in \cite[Lemma 2.2]{DemaineHIKP09}. Suppose there
	is a parent/child $(q,r)$ that heap property does not hold. Both
	$q,r$ cannot be in $\tau'_{t}$ by construction. The next touch time
	of elements outside $\tau_{t}'$ does not change, so both $q,r$ cannot
	be outside $\tau_{t}'$. 
	
	Now, we have $q\in\tau_{t}'$ and $r\notin\tau_{t}'$ where $N(r,t+1)<N(q,t+1)$.
	The rectangle defined from $(q,t)$ and $(r,N(r,t+1))$ will contradict
	Fact \ref{fact:sat on sides}. There are two sides to be considered.
	First, there is no point on the vertical side $((q,t),(q,N(r,t+1)]$
	because $N(r,t+1)<N(q,t+1)$. Next, all elements in $T_{t+1}$ between
	$q$ and $r$ must be descendants of $r$, and they cannot be touched
	as $r$ is not touched at time $t$. So the horizontal side $((q,t),(r,t)]$
	can only have one deletion point $s$ which has $q_{x}$ as a predecessor/successor.
	This violates Fact \ref{fact:sat on sides} and completes the proof.

\section{Proof Omitted from Section~\ref{sec:grdeq}}

\subsection{Proof of Lemma~\ref{lem:reduction concentrated}}

	Suppose that $S$ is not concentrated, and let $t_{0}$ be the first
	time when $S$ violates this condition. We modify the sequence
	and obtain another sequence $S'$ such that the first violation time
	is later than $t_{0}$, and the executions of $S$ and $S'$ on any
	BST algorithm are the same, and repeat the argument.
	
	So, assume w.l.o.g. that the element $x$ is inserted as the minimum
	at time $t_{0}$. Since the condition is violated, $x<y$ for some
	$y\in L_{t}$. Let $x'$ be an element such that $y<x'$,
	for all $y\in L_{t_{0}}$, and $x'$ is less than all elements in
	the current tree $T_{t_{0}}$. Note that $x'$ must exist,
	because there is no violation before time $t_{0}$.
	
	Now, since BST is a comparison-based model, as long as the relative
	values of all following update elements are preserved, even when the
	sequence is modified, the BST algorithm would behave the same.
	
	Therefore, we modify $S$ such that we set the value of $x$
	to be $x'$ while preserving the relative values of all following
	update elements. So now the condition is not violated at time $t_{0}$
	while the execution of the modified sequence is unchanged.

\end{document}